# ORIGINAL MINIHALO

**gravity confined by CDM spherical geometry**

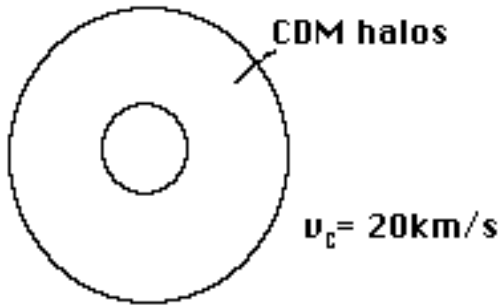

CDM halos
$v_c = 20$ km/s

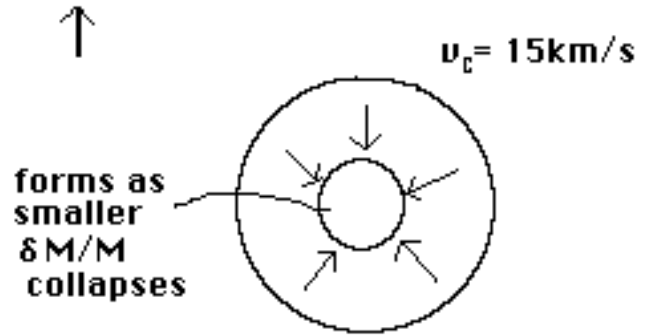

$v_c = 15$ km/s
forms as smaller $\delta M/M$ collapses

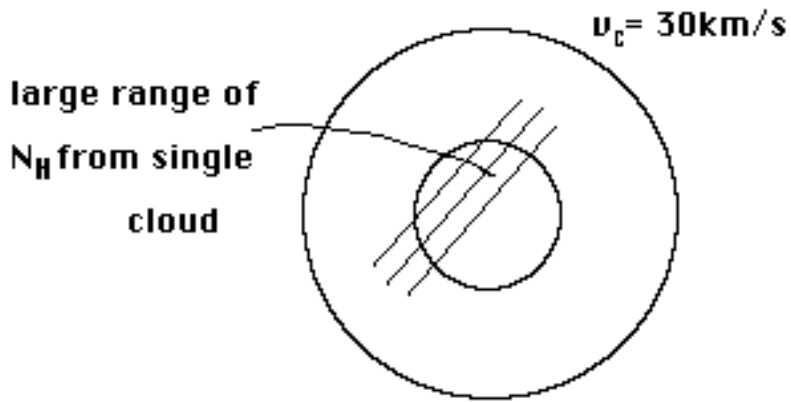

$v_c = 30$ km/s
large range of $N_H$ from single cloud

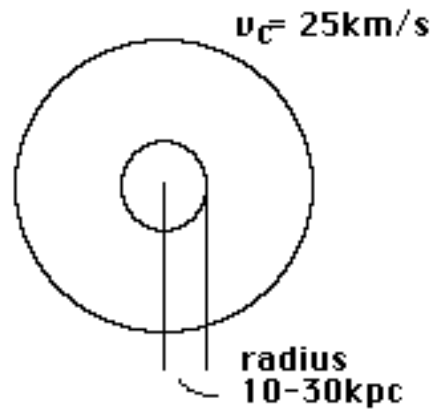

$v_c = 25$ km/s
radius 10-30 kpc

## ORIGINAL PRESSURE CONFINED

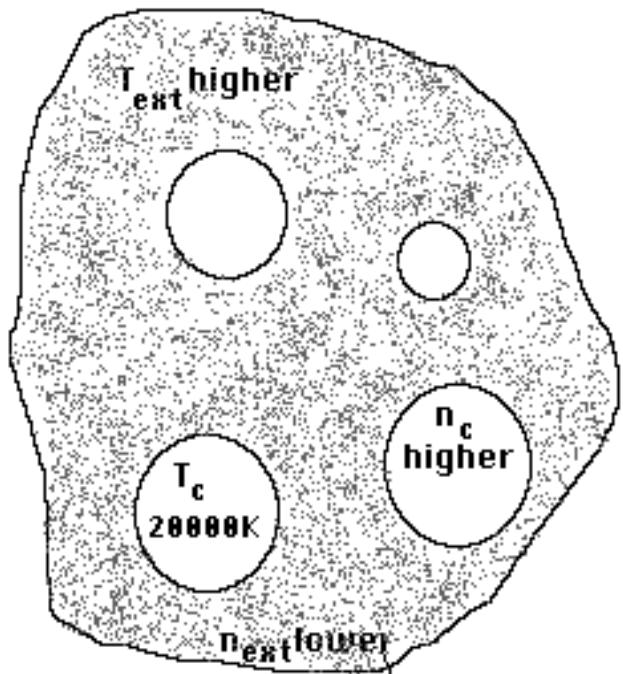

pressure can be uniform or local to clouds

## spherical "2 phases" in hydrostatic equil.

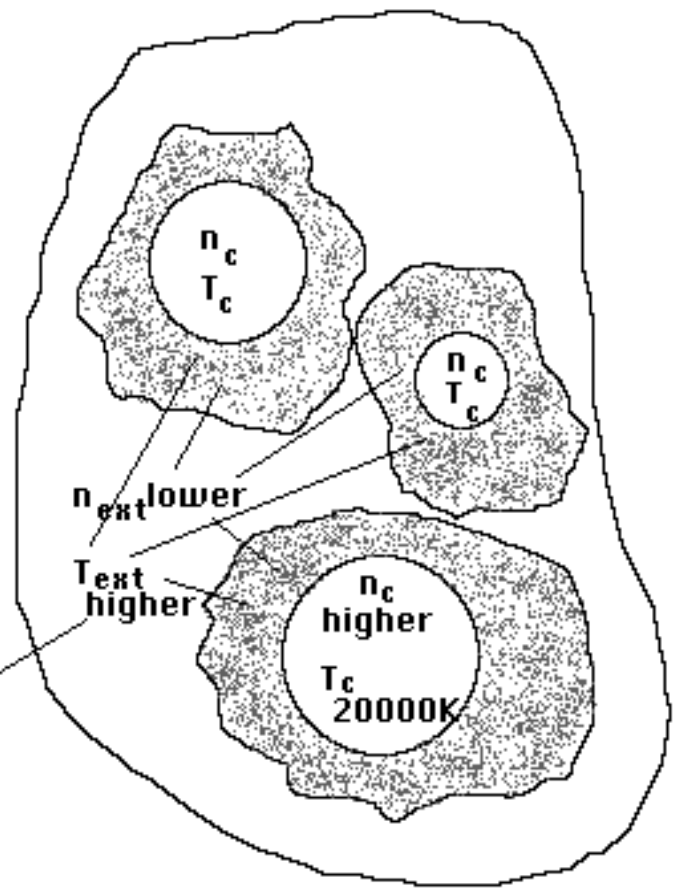

## PRESSURE CONFINED SLABS

gas can collapse into disk without forming stars

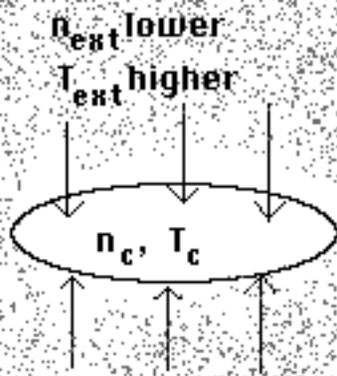

$n_{ext}$ lower
$T_{ext}$ higher

$n_c, T_c$

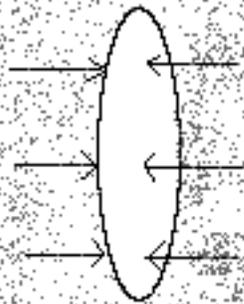

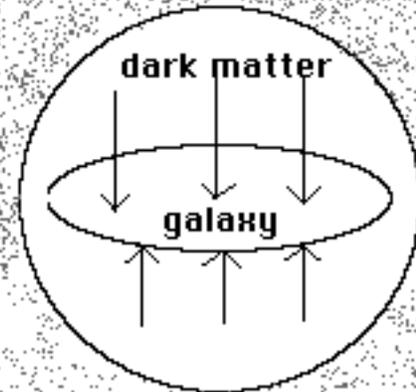

dark matter

galaxy

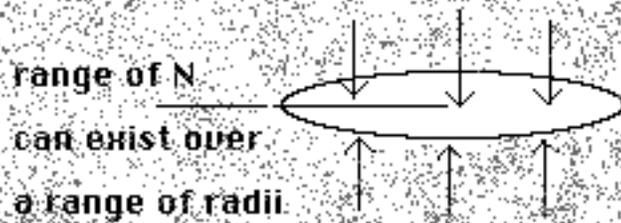

range of N can exist over a range of radii

## GRAVITY CONFINED SLABS   confined by CDM halos

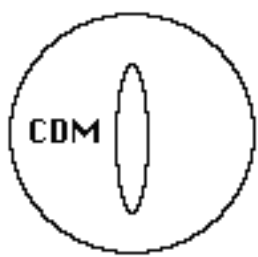
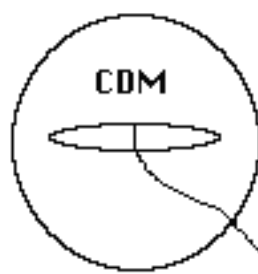
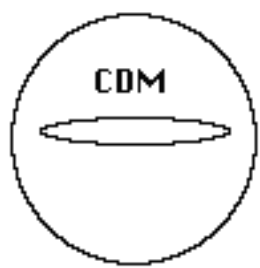

forest clouds have smaller central $N_H$ than galaxies

## MIXED PRESSURE AND GRAVITY - SLAB GEOMETRY

**example: Vanishing Cheshire Cats**

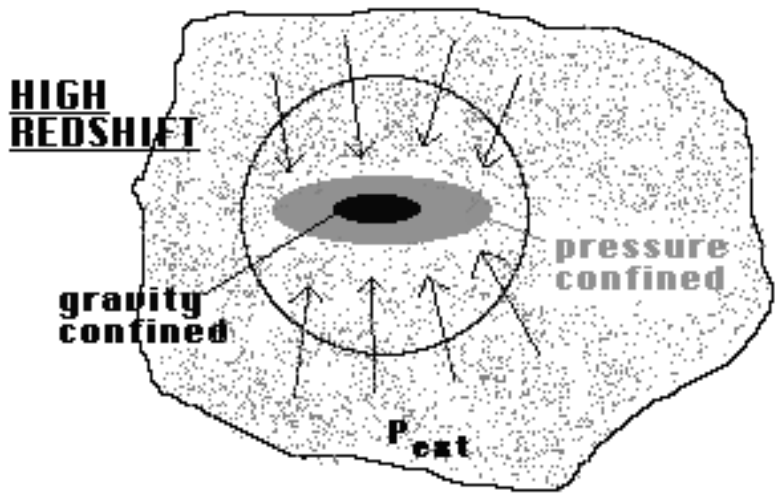

**HIGH REDSHIFT** — gravity confined, pressure confined, $P_{ext}$

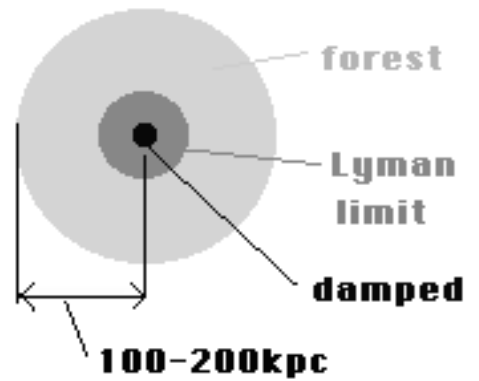

face on view — forest, Lyman limit, damped, 100–200 kpc

---

**FAINT BLUE GALAXY EPOCH** — stars form at the center and blow out gas, $z \simeq 1$

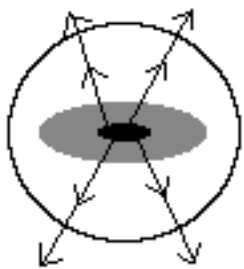

**PRESENT DAY** — outer disk could be pressure or gravity confined

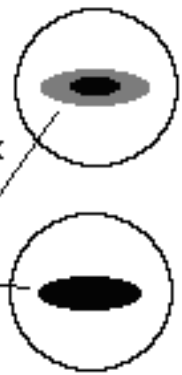

face on — forest, Lyman limit

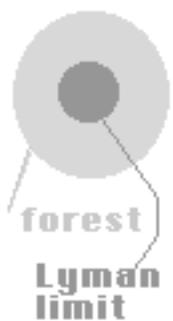

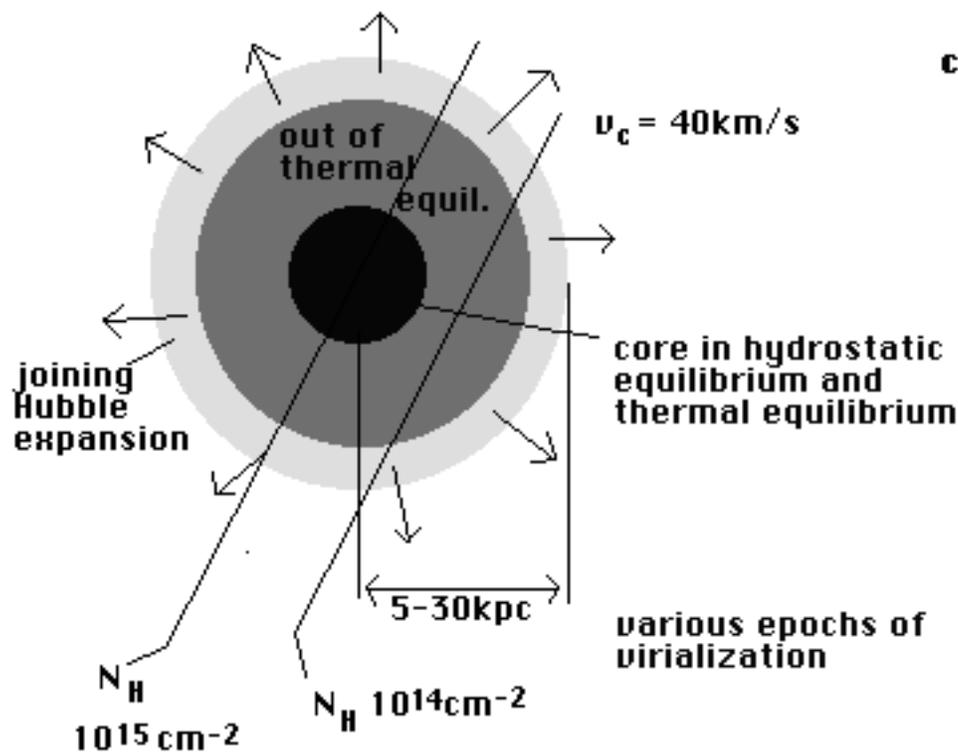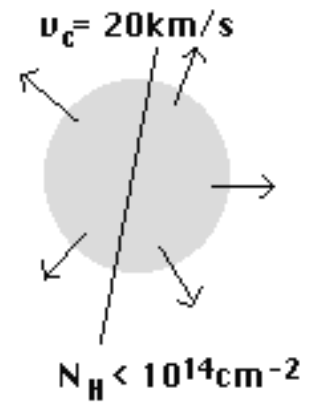

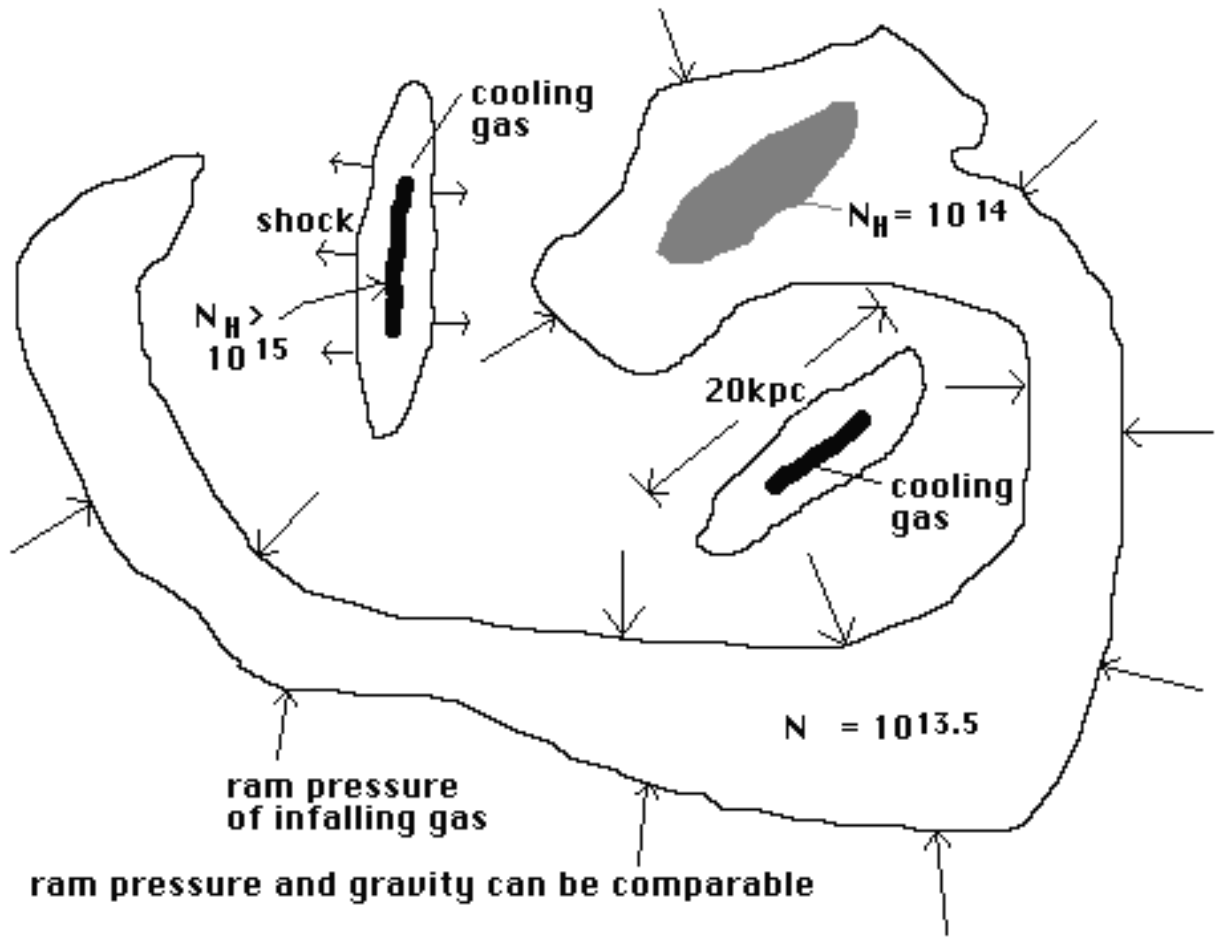

# Models of Lyα Forest Clouds

Jane C. Charlton

Astronomy and Astrophysics Department, Davey Laboratory, Pennsylvania State University, University Park, PA, 16802, USA

**Abstract.** The classic question regarding Lyα forest clouds is whether they are confined by the pressure of a hot, diffuse surrounding medium, or by the gravity of dark matter mini-halos. This paper reviews these basic models for forest clouds, considering spherical and slab geometries. At high redshift, the clouds are still likely to be in the process of formation, and it seems essential to consider them as dynamic structures in a cosmological context. At low redshift, observations of large cloud sizes (indicated by their covering of both lines of sight toward quasar pairs) have recently reshaped our view of the forest. Galaxy-like disks/slabs, which would be gravity confined near the center, but pressure confined in their outer regions (extending out to hundreds of kpcs), may be responsible for the low $z$ forest. The ultimate view of the identity of the forest clouds and their relationship to galaxies is likely to involve a synthesis of many of the models discussed here.

## 1 Introduction

A review of models of any type of object should naturally begin with a definition of the object. Yet, it is not simple to draw a picture of what we mean by a Lyα forest cloud. The picture is still being adjusted by knowledge of new data, and further, the clouds may not represent the same population over all time. In fact, it is likely that different experts in the field would have qualitatively distinct views of the types and distributions of the structures in which the Lyα clouds arise.

Why is it that the fundamental aspects of Lyα forest clouds have not yet been clarified? It is because knowing the number of clouds that intersect a given line of sight per unit redshift, $z$, and in some interval of neutral hydrogen column density, $N_H$, still allows many possibilities. A large density of small clouds or a few clouds that cover a large area could yield this same number per unit $z$. Also, different $N_H$ can be due to various lines of sight through the same object, or to different types (masses) of objects.

The picture that we would draw of a Lyα forest cloud is determined by our view of: 1) whether and how the cloud is confined (eg. by gravity or pressure), 2) how the cloud is formed, 3) the cloud geometry (eg. sphere, slab, or filament), 4) the relationship to galaxies. (Is it a separate population or does it in some way form a continuum with galaxies?) Observational constraints from lensed and double quasar lines of sight play a major role in shaping our view (Smette 1994; Bechtold et al. 1994, Dinshaw et al. 1994; Elowitz et al. 1994). By observing 4 forest clouds in common between lines of sight to the pair 0107-025A,B, Dinshaw et al. (1994; these proceedings) were able to place a model-independent lower limit of 160h$^{-1}$kpc on the cloud radii ($0.5 < z < 0.9$) assuming spherical geometries. Measurements of metals in the forest clouds (Tytler 1994, these proceedings) and detailed information on Doppler $b$ parameters and line profiles will also shape our view.

## 2    Pressure vs. Gravity; Sphere vs. Slab

This review begins by considering the four traditional models for Ly$\alpha$ forest clouds. These models consider pressure and gravity confinement, each in both a spherical and a slab geometry. For another perspective on models of forest clouds see the recent review by Bajtlik (1993).

### 2.1    Original Minihalo Model

In this model the clouds were taken to be spheres in hydrostatic equilibrium, confined by the gravity of a cold dark matter "mini-halo" (Rees 1986; Ikeuchi 1986; Ikeuchi et al. 1988). The confining halos were envisioned to have smaller masses than ordinary galaxy halos, with velocity dispersions of the order of tens of km/s. A "cartoon" view of this model is given in Fig. 1. In these spherical clouds, the density falls off rapidly with radius, so that it is possible to produce a large range of $N_H$ from a single cloud. This model has considerable predictive power as demonstrated in some recent papers (Miralda-Escude & Rees 1993; Mo et al. 1993). For current views on the mini-halo model see the article by Rees (1994) in these proceedings.

### 2.2    Original Pressure Confined Model

This model was discussed in the classic paper by Sargeant et al. (1980) and is illustrated in Fig. 2. Spherical clouds are confined in hydrostatic equilibrium by the pressure of a hotter, but more diffuse, exterior medium. The product of the density and the temperature of the cloud must balance with the external pressure. The pressure could be local to the clouds (in surrounding hot halos), or at the opposite extreme it could be a uniform intergalactic medium (IGM). This model has been developed by Ikeuchi and Ostriker (1986) and by Baron et al. (1989).

The pressure confined spherical cloud model has been found to have a serious problem. For a constant value of the external pressure, $P_{ext}$, the total number density $n_c$ of H (including both ionized and neutral) within the cloud is constant for all clouds. Thus the large range of $N_H$ for the forest cloud population ($10^{13}$ - $10^{17}$cm$^{-2}$) can only be a result of an unrealistically large range of cloud masses (Williger & Babul 1992). This can be demonstrated as follows.

For clouds in ionization/ recombination equilibrium we can write:

$$\alpha_{rec} n_c N_{tot} = \zeta N_H \qquad (1)$$

where $N_{tot}$ and $N_H$ are the total (ionized plus neutral) and neutral column densities, $\alpha_{rec}$ is the recombination coefficient, and $\zeta$ is the ionization rate due to the incident extragalactic background radiation. For a constant $n_c$ we have $N_{tot} \propto N_H$. Using $M_{tot} \propto N_{tot} R^2$ we find that $N_H \propto M_{tot}^{1/3}$. Thus a range of three orders of magnitude in $N_H$ requires a range of nine orders of magnitude in $M_{tot}$.

## 2.3 Pressure Confined Slabs

In a spiral galaxy we have a disk that has an outer region that contains gas, but no stars. Is it essential to the formation of a gaseous disk that stars will form at its center, and that it will become a galaxy? It is certainly not obvious that we could not form disks which have column densities below the threshhold for star formation. The resulting scheme for Lyα forest clouds is illustrated in Fig. 3. This raises the question of what fraction of the forest clouds are associated with ordinary galaxies. At low redshifts ∼ 50% of the forest clouds have a luminous galaxy within 160h$^{-1}$kpc (Lanzetta et al. 1994). However, at high redshift the larger numbers of forest clouds suggests that ordinary galaxies do not provide a sufficient cross section.

The model of pressure confined slabs was considered by Barcons and Fabian (1987). The slab geometry is advantageous in that it lends itself to large cloud dimensions, thus satisfying the observational size constraints mentioned in the introduction. The external pressure can be large enough to confine the clouds in the thin dimension, yet they can still be quite extended. However, the problem of requiring a huge range of cloud masses is merely rephrased as the problem of needing a variation of nine orders of magnitude in $N_{tot}$ over the radius of the slab. Furthermore, confining forest clouds with $N_H > 10^{16}$cm$^{-2}$ would require pressures that are likely to exceed constraints placed by the COBE y-parameter limits (inverse Compton scattering of microwave background photons off of the hot electrons) (Mather et al. 1994).

## 2.4 Gravity Confined Slabs

For completeness, we mention a fourth simple model, illustrated in Fig. 4, in which the mini-halos confine gas in a slab geometry. The velocity dispersions of these halos would still be less than those of ordinary galaxies, and the central gas column densities would be smaller. Large cloud dimensions are inconsistent with this simple class of models.

## 3    Mixed Pressure and Gravity Confinement (Slab Geometry)

It is possible, and in fact natural, that more than one mechanism is responsible for cloud confinement. Here we shall discuss a particular class of models in which forest clouds are produced by a single type of structure, a slab or disk which has its inner regions (with higher $N_H$) gravity confined and its outer regions (with smaller $N_H$) pressure confined. In this way it is possible to have very large disk sizes without an unreasonable range of $N_{tot}$ over the radius of the disk.

A particular example of this class of models has been discussed by Salpeter (1993) and further developed in Hoffman et al. (1993), Salpeter (1994), and Salpeter and Hoffman (1995). In this example, a hypothetical class of galaxies is responsible for producing the majority of the Ly$\alpha$ forest clouds. The idea is that the disks of these "Vanishing Cheshire Cat" galaxies have central column densities somewhat smaller than those of ordinary galaxies (perhaps a factor of ten smaller). In order to produce the observed covering factor of forest clouds at high redshifts these systems must be more abundant than ordinary galaxies (at least by a factor of 10). The column density would be exponential near the disk center, but would switch to a power law form further out, which could continue out to radii of a couple of hundred kpc. (In the outer disk regions it may be best to think of a collection of high velocity clouds as opposed to a smooth disk (Salpeter 1994).) This idea is partly motivated by 21cm observations of extended disks in dwarf galaxies (Hoffman et al. 1993). The large cloud radii that were envisioned in this class of models are quite consistent with the observed numbers of coincidences and anti-coincidences of Ly$\alpha$ forest lines along adjacent lines of sight toward double quasars These observations give a lower limit of $70h^{-1}$kpc at $z \sim 2$ down to a column density of approximately $10^{14.5}$cm$^{-2}$ (Bechtold et al. 1994; Dinshaw et al. 1994). Even larger sizes are likely since the clouds will probably extend to smaller $N_H$ at larger distances from cloud center. Most classes of models have difficulties producing such large cloud sizes (Bechtold et al. 1994; Dinshaw et al. 1994).

The evolution of the typical "Vanishing Cheshire Cat" is illustrated in Fig. 5. At high redshifts the inner portions of the disks could produce many of the lower column density damped systems (the highest columns would only come from ordinary galaxies), slightly larger radii would produce Lyman limit systems, and the outer disk would be responsible for the forest clouds. Star formation would be delayed until intermediate redshifts in these lower column density disks, and the first generation of stars would be responsible for ejecting the majority of the gas (as in Babul and Rees (1992)). The damped systems, and some of the Lyman limit systems contributed by this population, would disappear and the outer disk (the smile of the cat) would be left behind today as forest clouds. Observations of low redshift Lyman limit systems do suggest that there are relatively fewer with $N_H > 10^{18}$cm$^{-2}$ (as compared to the number in the range $10^{17} < N_H < 10^{18} \text{cm}^{-2}$) at recent times (Storrie-Lombardi et al. 1994). The inner region would be best described as a red population of very low surface brightness galaxies at the present time. With time, the transition radius between the gravity and pressure confined regions of the disk changes, subject to changes in the external confining pressure and the extragalactic ionizing radiation.

## 4   Pressure vs. Gravity

The value of the external pressure (whether it be local to the clouds or universal) determines the boundary between the inner, gravity confined and the outer, pressure confined regions of a slab. Balance of the forces in hydrostatic equilibrium can be written in the "half-slab approximation" (see Charlton et al. 1993; 1994) as:

$$\frac{\pi}{2} G m_H^2 \eta N_{tot}^2 + P_{ext} = 2 \, n_{tot} \, k \, T. \tag{2}$$

Here, $\eta$ is the contribution to gravity due to dark matter relative to that of ordinary gas. The gravitational and pressure forces are equally important at the specific value of

$$N_{tot} = N_1 = \left( \frac{2}{\pi} \frac{P_{ext}}{G m_H^2 \eta} \right)^{1/2}. \tag{3}$$

For $N_{tot} < N_1$ the slab is pressure confined, and using (1) we find that $N_H \propto N_{tot}$. In the gravity confined regime (for inner radii where $N_{tot} > N_1$) the specific relationship depends on the distribution of dark matter around clouds of different $N_{tot}$, but roughly $N_H \propto N_{tot}^3$.

The different relationships between $N_H$ and $N_{tot}$ translate to a prediction for the distribution function for the clouds (the number of clouds with a given column density). Specifically we know that the total column density distribution $g(N_{tot})$ is related to the neutral column density distribution $f(N_H)$ by:

$$g(N_{tot}) dN_{tot} = f(N_H) dN_H. \tag{4}$$

Each distribution is taken to have a power law form:

$$g(N_{tot}) \propto N_{tot}^{-\beta} \tag{5}$$

$$f(N_H) \propto N_H^{-\epsilon}. \tag{6}$$

Then we find, from the relationships between $N_H$ and $N_{tot}$ in the two regimes, that:

$$\beta = \epsilon \quad \text{pressure confined}$$
$$\beta = \frac{\epsilon + 2}{3} \quad \text{gravity confined.} \tag{7}$$

For $\epsilon = 1.8$, we find that $\beta = 1.8$ in the pressure dominated regime and $\beta = 1.3$ in the gravity dominated regime. This type of shape for $f(N_H)$ is apparent in the data of Petitjean et al. (1993a) (although some other studies do not agree (Kulkarni et al. 1994; Meiksin & Madau 1993)). If we assume the change in slope of $f(N_H)$ to be due to this transition, the value of $N_H$ at which the turnover occurs can be used to derive a value of the external pressure. In terms of the observed $N_H$ value, $N_{H1obs}$, combining (1) and (3), the pressure is given by:

$$\frac{P_{ext}}{k} = 5.1\eta_1^{1/3}\left(\frac{\zeta}{2.7\times 10^{-12}\mathrm{s}^{-1}} \times \frac{N_{H1obs}}{2\times 10^{15}\mathrm{cm}^{-2}}\right)^{2/3}\mathrm{cm}^{-3}\mathrm{K} \qquad (8)$$

The parameter $\eta_1$ relates to the ratio of dark matter to gas, and is expected to be of order 10. Thus at a redshift of 2.5 (where most of the Ly$\alpha$ forest data exists) the value is approximately $P_{ext}/k = 10\mathrm{cm}^{-3}\mathrm{K}$.

The specific non-power law shape of the observed $f(N_H)$ distribution was predicted by the transition from gas pressure confinement to dark matter gravity confinement, however a more general fact has also been demonstrated. If the mechanism of confinement changes at some column density, the distribution function of the clouds should also have a break at that column density. As more lines of sight are observed with high resolution (HIRES) by the Keck telescope it will be possible to accurately determine $f(N_H)$ and to chart its evolution with time.

It should be noted that this explanation of deviations from a power law distribution of the numbers of forest clouds is not unique. Petitjean et al. (1993a) have proposed that the change in shape at $N_H$ of $10^{15}$ or $10^{16}\mathrm{cm}^{-2}$ is due to a transition between metal-poor and metal rich systems. This is based on a detailed study of photoionization models of pressure confined, spherical clouds, with density profiles determined by gravity. The Petitjean et al. (1993b) models are an example of another class: a spherical geometry, including external pressure and gravity.

## 5   Cloud Structure and Formation Models

It seems likely that the simple models discussed above are in several ways unrealistic. Meiksin (1994) has performed one-dimensional hydrodynamic calculations in which he solves for the cloud structure in the context of the mini-halo model (in both spherical and slab geometry). These calculations do not assume the clouds to be in hydrostatic or in thermal equilibrium. Also, Meiksin consider the fact that the clouds are in a cosmological setting, i.e. matter is constantly accreting from outside the cloud. Murikama and Ikeuchi (1993) performed similar calculations, but did not find the same layered structure because their simulations were not performed in a cosmological context. As illustrated in Fig. 6, the larger clouds have an extent of 30kpc in radius, and they have a three phase structure. There is an inner core in hydrostatic and thermal equilibrium, a transition layer that is not in thermal equilibrium, and an outer accretion layer that joins onto the Hubble expansion. The core region is responsible for the higher column density forest clouds ($N_H$ of $10^{14}$ or $10^{15} \text{cm}^{-2}$), and the outer regions for those with lower column densities.

It has recently become possible to form Lyα forest clouds in the context of specific structure formation models using a "shock capturing" cosmological hydrocode (Cen et al. 1994). A specific application with initial conditions of cold dark matter and a cosmological constant is described by Miralda-Escude (these proceedings) so it will not be covered in detail here. However, the basic features are worth noting, because the picture differs substantially from the previous sketches. A schematic diagram is given in Fig. 7. The structures that would give rise to the lower $N_H$ forest clouds ($10^{14} \text{cm}^{-2}$) are often part of a filamentary, sheetlike network. Within the sheets there can be larger density enhancements that produce somewhat larger column densities. The sheet-like clouds are confined roughly equally by gravity and by the ram pressure of infalling gas.

## 6  The Death of Pressure Confinement is Premature

The predictive power and the successes in matching observations (Miralda-Escude & Rees 1993; Mo et al. 1993) seems to have led to a focus on gravity confined as opposed to pressure confined models. As pointed out by Rees (these proceedings) this eliminates the need for a hot intergalactic medium. On the other hand, the large sizes of clouds imply that another mechanism is needed in addition to gravity. It is not all that implausible that the intergalactic medium could be dense and hot enough to affect confinement of the low $N_H$ end of the forest.

In particular, it is worth noting that feedback from galaxy formation can have a substantial effect on heating the IGM (Cen & Ostriker 1993; Shapiro, Giroux, & Babul 1994). In fact the pressures produced by these cold dark matter based studies are large enough to dominate gravity for the confinement of small $N_H$ clouds. The IGM produced by this mechanism will be complex and will be hotter near to galaxies than in the voids. It is plausible that $\Omega_{IGM} = .02$ and $T = 10^6$K in the vicinity of clouds, and that would confine the forest by pressure up to column densities of $N_H = 10^{15} \text{cm}^{-2}$.

Alternatively, the confining pressure for large slab-like forest clouds could be due to hot gaseous halos that surround these clouds. In this case the confining pressure may increase with time. Alternatively, if the IGM was heated at a high redshift, then adiabatic cooling will lead to a decrease of the pressure with time.

## 7   Evolution of Lyα Clouds

The number of clouds as a function of $N_H$ and $z$ can in principle be used to constrain the models. The latest measurements of the rate of change of the ionizing background radiation $J$ and of the evolution of the number of clouds were described in Bechtold's contribution to these proceedings. In that contribution, having a better knowledge of $J$ allowed a better constraint on the simple equilibrium models. The goal of this section is not to provide a detailed interpretation of that data, but rather to explain the effect that numerous factors will have on cloud evolution. Let us quantify the evolution of clouds, with an equivalent width greater than some limiting value, as $N(z) \propto (1+z)^\Gamma$.

The ionizing UV background (likely due to quasars and young galaxies) decreases over the time that we observe the forest clouds. This pushes each cloud to a higher value of $N_H$ and thus (since there are more lower $N_H$ clouds than higher) will lead to more clouds as time goes by ($\Gamma$ decreases).

The external pressure could decrease adiabatically with time, or it could increase due to feedback from galaxies. A decreasing pressure results in fewer clouds confined by pressure, but the ones that are will have smaller densities $n_{tot}$ due to expansion. This results in a larger decrease in the number of recombinations relative to the decrease in number of ionizations (see (1)). Thus $N_H$ decreases for each pressure confined cloud and there are fewer clouds (a larger $\Gamma$) as time goes by.

Several other factors also affect evolution, and each is very poorly known. If clouds are expanding freely (not in equilibrium) it is expected that this effect would destroy the small clouds. Thus for those clouds $\Gamma$ would increase. Mergers will provide a shift in the distribution from more smaller clouds to more larger clouds. If small clouds are not constantly formed to replenish the supply this would lead to an increase in $\Gamma$ as well. Finally, if the number of new clouds that form decreases with time, then obviously $\Gamma$ can increase for that reason. It will undoubtedly be quite difficult to sort out all these effects.

The decrease in the ionization rate due to the extragalactic background radiation is expected to affect the observed sizes of individual clouds. Assume that the total column density distribution as a function of radius in a single cloud, $N_{tot}(R)$, does not change with time. As time goes on, a given $N_{tot}$ will correspond to a larger value of $N_H$, i.e. the cloud becomes more neutral. This also means that a given $N_H$ is found at a larger radius in the cloud at later times when the ionization rate has decreased. This is general conclusion for any model if $\zeta$ is decreasing. For a given detection threshold of column density (equivalent width), the apparent radius of a forest cloud, detected by its neutral Hydrogen content, is expected to increase with time. This is consistent with constraints on forest cloud sizes from quasar double lines of sight (Dinshaw et al. 1994; Bechtold et al. 1994; Dinshaw (these proceedings)).

## 8   Which Came First?

In closing, I'd like to pose a question that provides a reminder of how little we know for certain about the formation and evolution of the forest. The question is: which comes first, the galaxy, the Lyα clouds, or the IGM? Of course, one might say that, by definition, the IGM must have come first. However, in this case, let us consider the majority of the gas in the IGM, whether it was primordial or reproduced by galaxies, and when and if it was heated.

In fact, it is possible to envision three perfectly reasonable scenarios that produce the three possible orderings of these events. For example, in a simple, hierarchical structure formation model there was first an IGM, it clumped together small scales to form Lyα forest clouds, and these gradually merged together to make larger structures. Alternatively, the IGM could still be primarily leftover from galaxy formation, but Jean's mass constraints could lead to galaxy formation, followed by later fragmentation to form forest clouds, or ejection of them as proposed by Wang (1994). Finally, perhaps galaxies formed first and efficiently, and produced a considerable, moderately hot IGM by hydrodynamic feedback. This IGM may have been necessary to confine the Lyα forest clouds so that they were the last structures to form.

Surely, many other general scenarios can be proposed. It is clear that no single, simple model provides an adequate description of the Lyα forest. However, a combination of these ideas does seem to have a ring of truth. At high redshifts it is almost certainly the case that the forest clouds are dynamic structures, best explored by hydrodynamic models such as those of Cen et al. (1994). However, at low redshifts they are astoundingly large, coherent structures. Such structures seem to require some combination of gravity and pressure confinement. In addition, if the forest clouds were produced by objects with a radius of 100kpc, at a redshift of three or four they would have to be nearly overlapping to produce the observed covering factor. Some aspects of the simple models may be relevant for describing the low redshift structure and evolution of the forest clouds. Undoubtedly, the theoretical models that are shaped by these simple ideas will soon have overwhelming amounts of new data to confront.

*Acknowledgements.* This work was supported by NASA grant NAGW-3571. The ideas presented have been shaped by collaborations with Edwin Salpeter, Suzanne Linder, and Craig Hogan. I'd also like to thank Chris Churchill for his insightful criticisms of this manuscript.

# References

Babul, A., Rees, M.J., 1992, MNRAS, 255, 346.
Bajtlik, S., 1993, in The Environment and Evolution of Galaxies, eds. Shull, J. M. & Thronson, H. A., (The Netherlands: Kluwer), p. 191.
Barcons, X., Fabian, A.C., 1987, MNRAS, 224, 674.
Baron, E., Carswell, R.F., Hogan, C.J., Weymann, R.J., 1989, ApJ, 337, 609.
Bechtold, J., Crotts, A.P.S., Duncan, R.C., Fang, Y., 1994, ApJ, 437, L83.
Cen, R., Miralda-Escudé, J., Ostriker, J.P., Rauch, M. 1995, ApJ, in press.
Cen, R., Ostriker, J.P., 1993, ApJ, 417, 404.
Charlton, J.C., Salpeter, E.E., Hogan, C.J., 1993, ApJ, 402, 493.
Charlton, J.C., Salpeter, E.E., Linder, S.M. 1994, ApJ, 430, L29.
Dinshaw, N., Impey, C.D., Foltz, C. B., Weymann, R. J., Chaffee, F. H., 1994, ApJ, 437, L87.
Elowitz, R.M., Green, R.F., Impey, C.D., 1995, ApJ, in press.
Hoffman, G.L., Lu, N.Y., Salpeter, E.E., et al., 1993, AJ, 106, 39.
Ikeuchi, S., 1986, As&SS, 118, 509.
Ikeuchi, S., Murikama, I., Rees, M.J., 1988, MNRAS, 236, 21p.
Ikeuchi, S., Ostriker, J.P., 1986, ApJ, 337, 609.
Kulkarni, V.P., Huang, K.-L., Green, R. F., et al., 1994, ApJ, submitted.
Lanzetta, K.M., Bower, D.V., Tytler, D., Webb, J.K., 1995, ApJ, in press.
Mather, J. C. et al., 1994, ApJ, 420, 439.
Meiksin, A., 1994, ApJ, 431,109.
Meiksin, A., Madau, P., 1993, ApJ, 412, 34.
Miralda-Escudé, J., Rees, M.J., 1993, MNRAS, 260, 617.
Mo, H.J., Miralda-Escudé, J., Rees, M.J. 1993, MNRAS, 264, 705.
Murikama, I., Ikeuchi, S., 1993, ApJ, 409, 42.
Petitjean, P., Webb, J.K., Rauch, M., et al., 1993a, MNRAS, 262, 499.
Petitjean, P., Bergeron, J., Carswell, R.F., Puget, J.L., 1993b, MNRAS, 260, 67.
Rees, M.J., 1986, MNRAS, 218, 25p.
Salpeter, E.E., 1993, AJ, 106, 1265.
Salpeter, E.E., 1994, preprint.
Salpeter, E.E., Hoffman, G.J., 1995, ApJ, in press.
Sargeant, W.L.W., Young, P.J., Boksenberg, A., Tytler, D., 1980, ApJSupp, 42, 41.
Shapiro, P.R., Giroux, M.L., Babul, A., 1995, ApJ, in press.
Smette, A., Surdej, J., Shaver, P.A. et al., 1992, ApJ, 389, 39.
Storrie-Lombardi, J. L., McMahon, R. G., Irwin, M. J., Hazard, C., 1994, ApJ, 427, L13.
Wang, B., 1995, ApJ Letters, in press.
Williger, G. M., & Babul, A., 1992, ApJ, 399, 385.